\documentstyle[twocolumn,epsf]{jpsj}

\newcommand{\I}{{\rm i}}
\newcommand{\EXP}{{\rm e}}
\newcommand{\nn}{\nonumber}

\title{Double Phase Transitions in Magnetized Spinor
Bose-Einstein Condensation}
\author{
  Tomoya {\sc Isoshima},
  Tetsuo {\sc Ohmi$^{1}$} and
  Kazushige {\sc Machida}
}
\inst{Department of Physics, Okayama University, Okayama 700-8530, 

$^{1}$Department of Physics, Graduate School of Science, Kyoto University, Kyoto 606-8502}
\recdate{\today}
\abst{
It is investigated theoretically that magnetized Bose-Einstein condensation (BEC) with the internal (spin) degrees of freedom exhibits a rich variety of phase transitions, depending on the sign of the interaction in the spin channel.
In the antiferromagnetic interaction case there exist always double BEC transitions from single component BEC to multiple component BEC. 
In the ferromagnetic case BEC becomes always unstable at a lower temperature, leading to a phase separation.
The detailed phase diagram for the temperature vs the polarization, the spatial spin structure, the distribution of non-condensates and
the excitation spectrum are examined for the harmonically
trapped systems.
}
\kword{
Bose-Einstein condensation, Non-condensate,
Gross-Pitaevskii equation, Bogoliubov equation,
Critical Temperature
}

\begin{document}
\sloppy
\maketitle

\section{Introduction}

There has been much attention focused on
the Bose-Einstein condensation of
alkali-metal atom gases both 
experimentally and theoretically recent years
since the experimental realizations in 
1995~\cite{firstRb,hulet,ketterle}.
Interest ranges widely from atomic physics, 
quantum optics 
to many-body physics.
One of the focused points from the many-body
physics viewpoint lies in the fact that 
the Bose-Einstein condensation
with internal degrees of freedom can be realized, thus allowing us 
to investigate the interplay between superfluidity and 
internal degrees of freedom.
Namely, in the recent experiments on Na atoms
via an optical trapping method
the three hyperfine states with $m_F=1, 0, -1$ of $F=1$
for the case of Na simultaneously are Bose condensed
as named a spinor BEC~\cite{stamper}.
This contrasts with much widely used magnetic trapping
experiments where the freedom due to various hyperfine states is
frozen and BEC is described by a scalar order parameter.

Prior to the realization~\cite{stenger} of a spinor BEC,
only a few examples of the superfluid with internal degrees of freedom
such as spin or orbital degrees of freedom are 
known like in a neutral Fermion system: $^3$He~\cite{woelfle} or
a charged Fermion system: UPt$_3$~\cite{machidaUPt3}.
An advantage for considering dilute alkali-metal atom gases
lies in the fact that these are weakly interacting Bose systems,
thus mean field treatment of Bogoliubov theory is
well applicable, which is a controllable approximation.
This feature is absent in $^3$He and UPt$_3$.
This allows us to construct a theory from a
microscopic Hamiltonian without any adjustable parameter.
This is one of unique properties in the present alkali-metal BEC systems.

In this paper,
we consider the phase diagram of the BEC with internal degrees of freedom
characterized by having the atomic hyperfine
state with $F=1$. Thus the triple degenerate components
($m_F=1,0,-1$) are available and can simultaneously Bose condense
by an optical trapping.
It is well known that the condensation temperature
decreases when the spin degrees of freedom exist
than when the spin degree of freedom does not exist,
namely the degeneracy lowers the BEC critical temperature.
Therefore a magnetically trapped system has generally
a higher BEC transition temperature than an optically trapped system,
if the other conditions are same.

On the other hand, the total magnetization of the system is preserved
in BEC because the system is isolated
and the magnetic dipole interaction is weak.
In this case with $m_F=1,0,-1$,
the two phase transitions are expected rather than three
when the system has a finite magnetization. 
The majority spin component, say, $m_F=1$ whose density
is larger than the others first Bose condenses upon lowering temperature $T$.
The second transition occur
only when all three chemical potentials, which were
spaced equally by the Zeeman effect in the normal state, coincide.
Thus the third transition is never realized for a finite magnetization from the general ground.
This double transition phenomenon is similar to
the split transitions to the A1 phase and A2 phase 
of superfluid $^3$He-A under the magnetic field.~\cite{woelfle}
It is an interesting problem how these transition temperatures actually
varies when  taking into account the polarization as a controlling parameter of the system.
It is also interesting to see how the interaction makes
the nature of the phase transition alter and the critical
temperatures change relative to ideal Bose systems.
The spin dependent
interaction constant which characterizes the spinor BEC
could be either ferromagnetic and antiferromangentic.
The former is anticipated in $^{87}$Rb~\cite{stamper}
while the latter is realized in $^{23}$Na with $F=1$. 
We will see that the sign of this interaction channel
is crucial in determining the low temperature properties
even though its strength may be small in actual systems.


In the next section, the phase diagram in a system of non-interacting Bosons
is discussed as a preliminary.
A microscopic formulation based on a mean-field theory
of Bogoliubov approximation is introduced and
the Popov equations which describe interacting Bosons with the
spin degrees of freedom
are derived in Sec.~\ref{sec:popov}, following
our series of papers~\cite{ohmi_bec,isoshima} and others~\cite{ho}
 on spinor BEC.
Using them, we discuss the phase diagram of the interacting Boson systems.
The effects of the interaction both for ferromagnetic and
antiferromagnetic cases are discussed there. 
The last section is devoted to summary and discussions.

\section{Ideal Bose systems}\label{sec:ideal}

In order to facilitate the later discussions on the interaction effects of the spinor BEC, we first treat a system of non-interacting particles with $F=1$.
The system is assumed to have a given polarization $M$ and cylindrical symmetry along the $z$ direction.
It is trapped harmonically with the trap frequency $\omega/2\pi$ in the radial direction.
The three kinds of the particle number $N$ of non-condensed atoms, which are specified by the hyperfine, or spin states $m_F=1, 0, -1$, are written as
\begin{equation}
N(\beta |\mu_i|) = \sum^\infty_{n_x,n_y=0}\int_{-\infty}^\infty
    \frac{dk_z}{\exp \beta(\varepsilon(n_x, n_y, k_z) +|\mu_i|) - 1},
\end{equation}
where the chemical potential $\mu_i (\leq 0)$
corresponds to each spin state ($i=1, 0, -1$) 
and they are split equally by the Zeeman effect:
$\mu_1+\mu_{-1}=2\mu_0$ to maintain the given polarization.
The eigenvalue is given by
\begin{equation}
\varepsilon(n_x, n_y, k_z) = \hbar \omega ( n_x + n_y )
    + \frac{\hbar^2 }{2m}(\frac{2\pi}{L})^2 k_z^2,
\end{equation}
$\beta = 1/k_B T$ and $L$ is the length along the $z$ direction.
$N(\beta \mu_i)$ is simplified to 
\begin{eqnarray}
N(\beta |\mu_i|) &=& C f(\beta \mu_i) T^{5/2} 
\\
C &=& \sqrt{\pi}k_B^{5/2} \frac{\sqrt{2m} L }{(\hbar\omega)^2 h},
\\
f(\beta |\mu_i|) &=&  \sum_{n=1}^{\infty} \frac{e^{- \beta |\mu_i| n}}{n^{5/2}}.
\end{eqnarray}

Let us consider the critical temperature $T_{c1}$,
which is determined by the condition $\mu_1=0$
where the condensate with the spin state $i=1$
becomes non-vanishing from high temperature.
Since $\mu_1+\mu_{-1}=2\mu_0$, the particle numbers of the spin state of $+1, 0$ and $-1$ are given by
$N(0), N(a)$ and $N(2a)$ respectively
where the dimensionless implicit parameter $a$ is proportional to
the Zeeman splitting.
The particle number $N$ and the polarization $M$
which are given parameters of the problem are
expressed in terms of the implicit parameter $a$ as
\begin{eqnarray}
  N &=& CT_{c1}^{5/2}\left\{ f(0)+f(a)+f(2a) \right\}, \label{eq:tc1:n}
\\
  M &=&  C T_{c1}^{5/2}\left\{ f(0) - f(2a) \right\}.  \label{eq:tc1:m}
\end{eqnarray}
When all particles in the system are in the state $i = +1$ and the fully polarized case: $M/N=1$,
we denote the critical temperature as $T_0$.
This corresponds to the limit where only the single component condenses,
that is, $a \rightarrow \infty$.
The total particle number at $T=T_0$ is
\begin{eqnarray}
  N =  C  T_0^{5/2}f(0).       \label{eq:tc1zero:n}
\end{eqnarray}
Comparing $N$ in eqs.\ (\ref{eq:tc1:n}) and (\ref{eq:tc1zero:n}),
we obtain 
\begin{equation}
  \frac{T_{c1}}{T_0} =
  \left[ \frac{f(0)}{f(0) + f(a) + f(2a)} \right]^{2/5}.
  \label{eq:tc1} 
\end{equation}
The relative polarization $M/N$ is derived from
eqs.\ (\ref{eq:tc1:n}) and (\ref{eq:tc1:m}) as
\begin{equation}
    \frac{M}{N} = \frac{f(0) - f(2a)}{f(0) + f(a) + f(2a)}.
   \label{eq:mn}
\end{equation}
Equations (\ref{eq:tc1}) and (\ref{eq:mn}) constitute the coupled equations to determine 
$T_{c1}/T_0$ as a function of $M/N$
through the parameter $a$.

Let us consider the second critical temperature $T_{c2}$ at which all three components simultaneously condense, namely, at which $a=0$.
The particle numbers of the spin components with
$+1, 0$ and $ -1$ are given by
$N_c + N(0), N(0)$ and $N(0)$ respectively,
where $N_c$ is the number of atoms in the condensate.
The total atom number and the polarization are expressed as
\begin{eqnarray}
  N &=& N_c + 3  C T_{c2}^{5/2} f(0),
\\
  M &=& N_c,
\end{eqnarray}
thus giving
\begin{equation}
N- M = 3 C T_{c2}^{5/2}f(0).   \label{eq:tc2:n_minus_m}
\end{equation}
We can consider the point which yields same polarization $M$ on $T_{c1}$ line as a function of $M$.
The value $N-M$ is evaluated from
eqs.\ (\ref{eq:tc1:n}) and (\ref{eq:tc1:m}) as
\begin{equation}
   N-M =C  T_{c1}^{5/2} \left\{ f(a) + 2 f(2a) \right\}.
                                           \label{eq:tc1:n_minus_m}
\end{equation}
Comparing eqs.\ (\ref{eq:tc2:n_minus_m}) and (\ref{eq:tc1:n_minus_m}),
we find
\begin{equation}
    \frac{T_{c2}}{T_{c1}} = \left( \frac{ f(a) + 2 f(2a) }{ 3 f(0) } \right)^{2/5} .
    \label{eq:tc1tc2}
\end{equation}

Equations (\ref{eq:tc1}) and (\ref{eq:tc1tc2}) determine
the $T_{c1} / T_0$ and $T_{c2} / T_0$
as a function of $M/N$.
It is noted that at $M/N=0$, $T_{c1}/T_0=3^{-2/5}=0.644$.
This is reasonable because the critical temperature decreases as the degeneracy of the system increases.
Note that the above factor 3 comes from the triple degeneracy.

\begin{figure}
   \begin{center}\leavevmode
   \epsfxsize=6cm \ \epsfbox{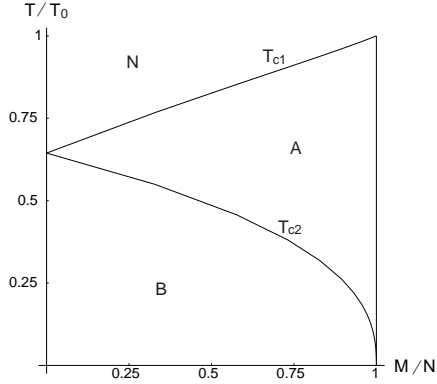} \\
   \end{center}
   \caption{
     The phase diagram of the non-interacting system.
     The axes are the normalized polarization $M/N$
     and the normalized temperature $T / T_0$.
     The A (B) phase is characterized by the single (three) spin component BEC. N denotes the normal state. 
   }
\label{fig:tc1tc2}
\end{figure}

Figure\ \ref{fig:tc1tc2} shows the
two critical temperatures
$T_{c1}$ and $T_{c2}$ as a function of the given  polarization.
It is found that at a fixed polarization $M/N(\neq 0)$
upon decreasing $T$ the system always shows the double transitions from the single spin component condensation; the A phase at $T_{c1}$
to the three component condensation; the B phase
at $T_{c2}$.
This double transition feature is similar to the A1 and A2 transitions in the A phase of the superfluid $^3$He under an external field.

\section{Formulation for interacting Bosons}\label{sec:popov}

In order to treat an interaction Boson system,
we give the formulation based on Bogoliubov theory, which is extended to take into account the spin degrees of freedom.
We start with a general 
Hamiltonian which fully considers the system's symmetries,
such as translation and rotations in real space and 
spin space:
\begin{eqnarray}
   H &=&
   \int\! d{\bf r} [
      \sum_{ij}\Psi_i^{\dagger}
          \left\{h({\bf r})\delta_{ij}-{\mathcal{B}}_{ij}\right\}
      \Psi_j
\nn\\&&\  %
      + \frac{g_n}{2} \sum_{ij}\Psi_i^{\dagger} \Psi_j^{\dagger} \Psi_j \Psi_i 
\nn\\&&\  %
      + \frac{g_s}{2} \sum_{\alpha}
          \sum_{ijkl} \Psi_i^{\dagger}\Psi_j^{\dagger}
          (F_{\alpha})_{ik}(F_{\alpha})_{jl}
          \Psi_k \Psi_l
      ],
\end{eqnarray}
where 
\begin{eqnarray}
   h({\bf r}) &=& - \frac{\hbar^2 \nabla^2}{2m} - \mu + V({\bf r}),
\\
  g_n &=& \frac{4 \pi \hbar^2}{m} \cdot \frac{a_0 + 2a_2}{3},
\\
  g_s &=& \frac{4 \pi \hbar^2}{m} \cdot \frac{a_2 - a_0}{3} .
\end{eqnarray}
The subscripts are $\alpha = (x,y,z)$ and $i,j,k,l = (0, \pm1)$.
The matrix $\mathcal{B}$ is a diagonal matrix whose elements are (1, 0, -1) and expresses the magnetic field
acting on the spins as the Zeeman effect. 
The harmonic confining potential is
$V({\bf r})={1\over 2}m\omega^2r^2$.

Following our previous paper,~\cite{ohmi_bec}
and the usual procedure for BEC system without spin,
the Gross-Pitaevskii equation including the non-condensate
as the mean field is obtained as
\begin{eqnarray}
   [
      h({\bf r}) \delta_{ij}  - {\mathcal{B}}_{ij}
&&\nn\\
      + g_n\left\{
         \sum_k \left( |\phi_k|^2 + \rho_{kk}\right) \delta_{ij} 
         +\rho_{ij}^{\ast}
      \right\}
&&\nn\\
      + g_s \sum_{\alpha}\sum_{kl}
      \{
         (F_{\alpha})_{ij}
         (F_{\alpha})_{kl} (\phi_k^{\ast} \phi_l + \rho_{kl})
&&\nn\\ \quad
         + (F_{\alpha})_{ik} \rho_{kl}^{\ast} (F_{\alpha})_{lj}
      \} ] \phi_j &=& 0.
\end{eqnarray}
The non-condensate density
\begin{equation}
   \rho_{kl} = \langle
      \hat{\psi_k}^{\dagger}({\bf r})
      \hat{\psi_l}({\bf r})
   \rangle
   \label{eq:rho}
\end{equation}
is introduced.

The non-condensate density is obtained through
the excitation spectrum, which is the solution of the
Popov equations given by
\begin{eqnarray}
   \sum_j \{
        A_{ij}     u_q({\bf r}, j)
      - B_{ij}     v_q({\bf r}, j)\} &=& \varepsilon_q u_q({\bf r}, i),
\\
   \sum_j \{
        B^{\ast}_{ij}   u_q({\bf r}, j)
      - A^{\ast}_{ij}   v_q({\bf r}, j)\} &=& \varepsilon_q v_q({\bf r}, i).
\end{eqnarray}
where
\begin{eqnarray}
   A_{ij} &=&
      h({\bf r}) \delta_{ij} - {\mathcal{B}}_{ij}
\nn\\&&
      + g_n \left\{
         \sum_k \left( |\phi_k|^2 + \rho_{kk} \right) \delta_{ij}
         +\left( \phi_i \phi_j^{\ast} + \rho_{ij}^{\ast} \right)
      \right\}
\nn\\&&
      + g_s
         \sum_{\alpha}\sum_{kl} [
            (F_{\alpha})_{ij}
            (F_{\alpha})_{kl} \left(
               \phi_k^{\ast} \phi_l + \rho_{kl}
            \right)
\nn\\&&\quad
            +(F_{\alpha})_{il}
            (F_{\alpha})_{kj} \left(
               \phi_k^{\ast} \phi_l + \rho_{kl}
            \right)
         ],
\label{eq:a}
\\
   B_{ij} &=&
      g_n \phi_i \phi_j
      +g_s 
         \sum_{\alpha} \sum_{kl} \left[
            (F_{\alpha})_{ik} \phi_k  (F_{\alpha})_{jl} \phi_l
         \right],
\label{eq:b}
\end{eqnarray}
$u_q({\bf r}, i)$ and $v_q({\bf r}, i)$ are the $q$-th eigenfunctions with the spin $i$ and
$\varepsilon_q$ corresponds to the $q$-th eigenvalue.

Equation\ (\ref{eq:rho}) is rewritten in terms of these eigenfunctions as
\begin{equation}
    \rho_{ij}({\bf r}) = \sum_q
       [u^{\ast} _q({\bf r}, i)u_q({\bf r}, j) f(\varepsilon_q)
      + v_q({\bf r}, i) v^{\ast} _q({\bf r}, j)( f(\varepsilon_q) + 1 )].
\end{equation}

We are treating a cylindrical system ${\bf r} = (r, \theta, z)$
and assume that
each Boson does not have the spatial angular momentum.
The relevant physical quantities are written in the
following form up to the phase factor:
\begin{eqnarray}
    \phi_j({\bf r}) &=& \phi_j(r),
\\
    u_q({\bf r}, j)    &=&
    u_q(r, j) \EXP^{\I q_{\theta}\theta} \EXP^{\I q_{z}z},
\\
     v_q({\bf r}, j)    &=&
    v_q(r, j) \EXP^{\I q_{\theta}\theta} \EXP^{\I q_{z}z},
\\
    \rho_{ij}({\bf r}) &=& \rho_{ij}(r).
\end{eqnarray}
Thus the system is specified by the quantum numbers;
the angular momentum  $q_{\theta}$ and the momentum $q_z$ along the $z$ direction.
The density of atoms in the spin state $i$ is $|\phi_{i}|^2 + \rho_{ii}$.
Therefore, the total particle number and the polarization
corresponding to eqs.~(\ref{eq:tc1:n}) and (\ref{eq:tc1:m}) are given by
\begin{eqnarray}
 N &=& \int \sum_i (|\phi_{i}|^2 + \rho_{ii}) d{\bf r}, \\
 M &=& \int \{ (|\phi_{1}|^2 + \rho_{11}) -
     (|\phi_{-1}|^2 + \rho_{-1 -1})\} d{\bf r}.
\end{eqnarray}
%
We have done the self-consistent calculations to solve
the above coupled equations.
The actual numerical computations are performed under
the conditions:
The mass and the scattering length of atoms are
$m = 3.81 \times 10^{-26} {\rm kg}$ and $a_0 = 2.75 {\rm nm}$.
These are equal to those of Na atom.
The other scattering length $a_2$ is defined so that $g_s$ becomes $\pm 0.1$.
The area density per unit length along the $z$ axis
is $N / L = 2 \times 10^4 (\mu m)^{-1}$.
The system size of $z$ direction is $L = 10 \mu m$.
The trapping frequency is $\omega / (2\pi) = 200$Hz.
The energy is scaled by the trap unit $\hbar \omega$.
The resulting transition temperature $T_0$ for the 
non-interacting Boson system is $T_0=0.757\mu K$.

The excitation levels whose energy exceeds  $50 \hbar \omega$ are
calculated by ignoring the interaction energy.
Most of the neglected energy comes form the  
diagonal part of the $g_n$ term in eqs.~(\ref{eq:a}) and (\ref{eq:b}).
That is approximately estimated as $g_n \times $ (density) $\simeq  7 \hbar \omega$.
This is sufficiently smaller than  the cutoff $50 \hbar \omega$,
meaning that the energy levels above the cutoff are not affected by the interactions and the above cutoff procedure is reasonable.


\section{Interacting Bose systems}\label{sec:interact}

By taking into account the presence of the non-condensates, we calculate a self-consistent set of the physical quantities under given parameters such as $g_n$, $\omega$,
$T$ or the polarization $M/N$. We assume that the fictitious magnetic field $B$ is applied to the system, which is treated as a Lagrange multiplier and fix the
polarization value. Both ferromagnetic ($g_s<0)$)
and antiferromagnetic  ($g_s>0)$)  cases are comparatively 
studied in the following. The results will be compared with those in the
previous non-interacting case. We see how the system is 
affected by the interaction, in particular its spin dependent channel in the light of the actual experimental situation for $^{23}$Na atom gases with $F=1$.

\begin{figure}
   \begin{center} \leavevmode

   \epsfxsize=6cm \ \epsfbox{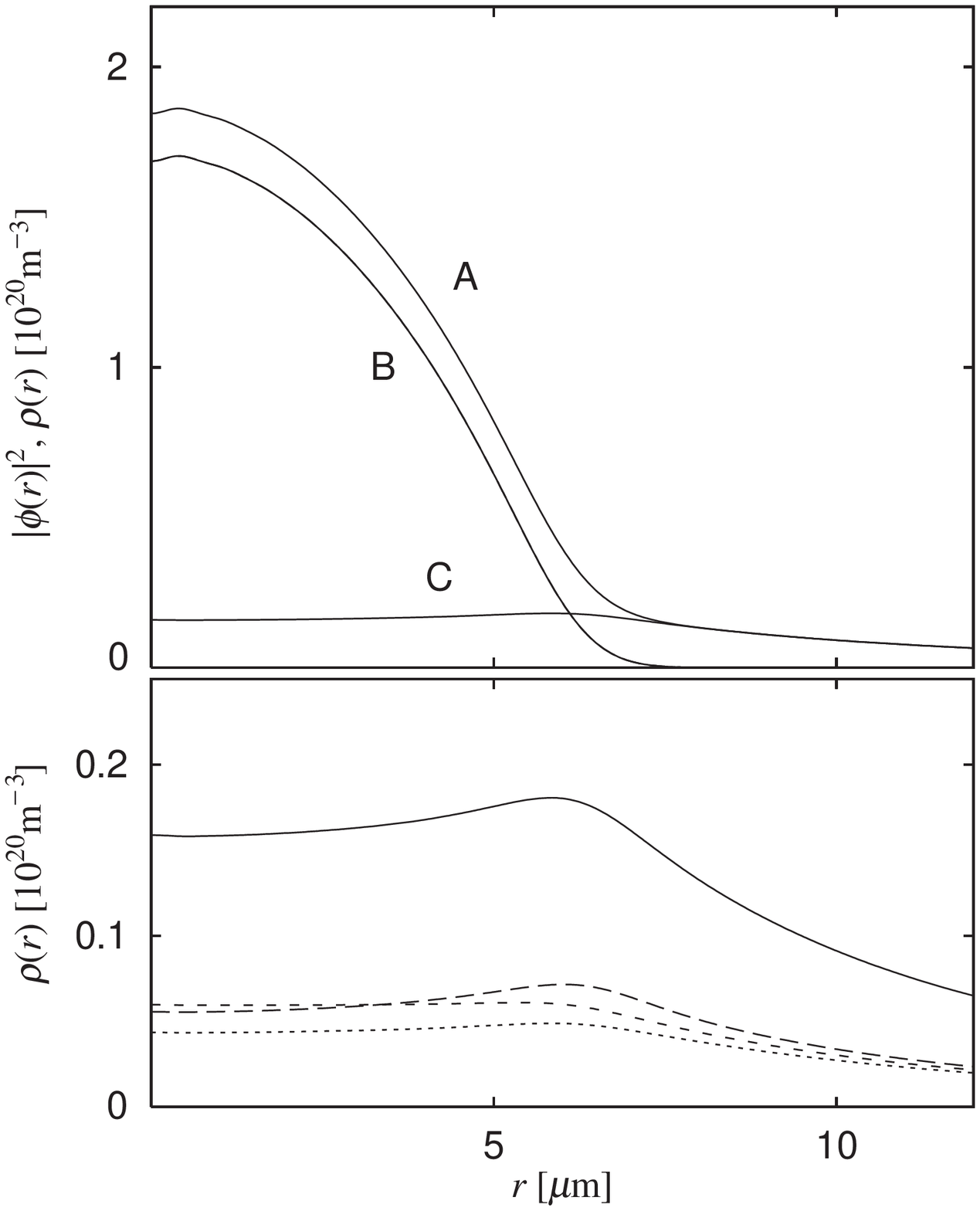} \\
   (a)\\

   \vspace{1\baselineskip}
   \epsfxsize=5cm \ \epsfbox{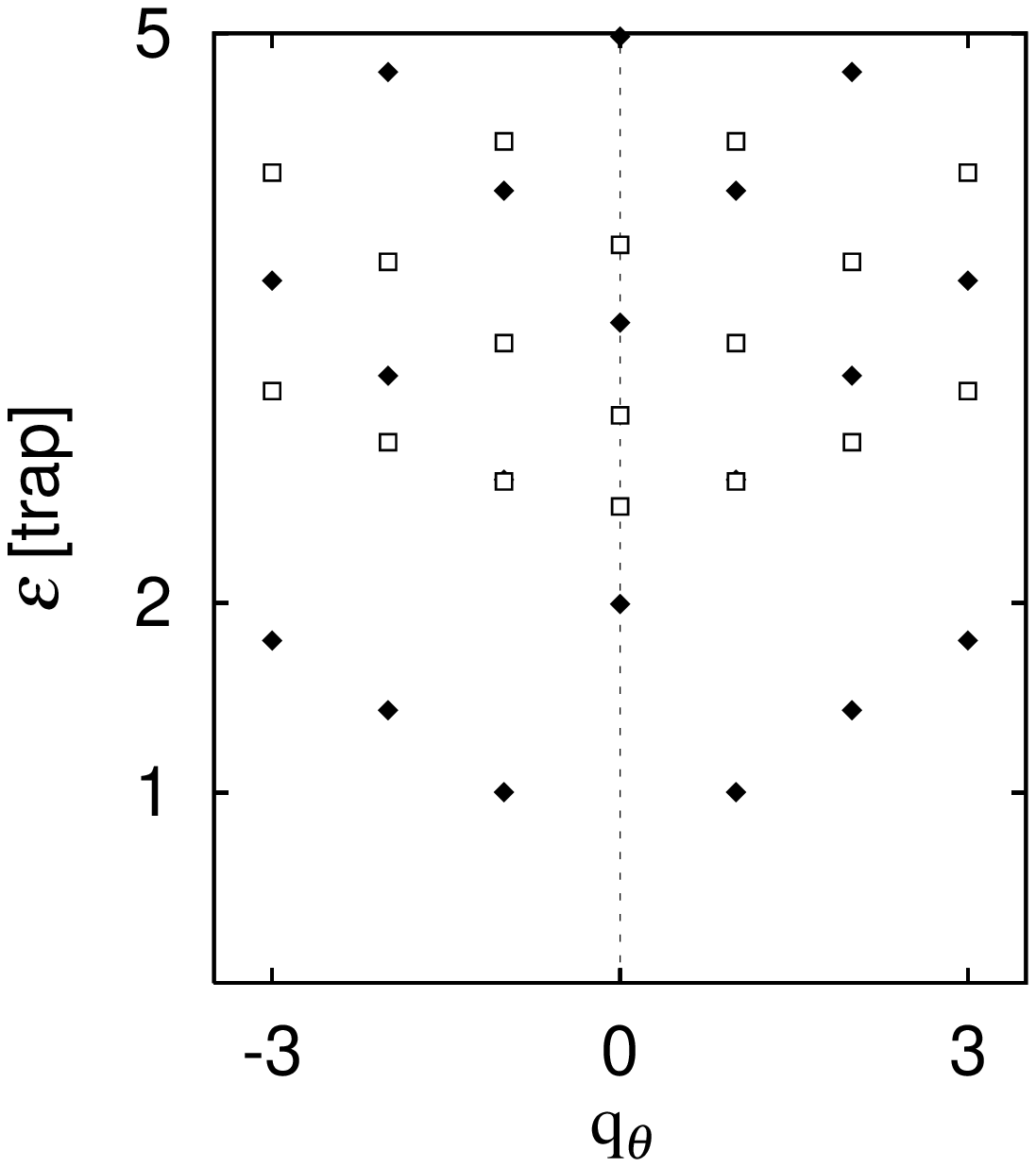} \\
   (b)\\

   \end{center}
   \caption{
   (a) Upper panel: Density profiles of the total number (A), the condensate with the spin +1 component (B) and the non-condensate (C). Lower panel: Enlarged density profiles of the non-condensate. The total (solid line), the $m_{\rm F} = +1$ component (long dashed line), 0 (short dashed line) and -1 (dotted line). 
   (b) Excitation level scheme in the lower energy region
as a function of the angular momentum $q_{\theta}$ at $q_z$=0
where $g_s/g_n = -0.1$, $T = 0.3 \mu K$
and the polarization $M/N = 0.586$.
The filled dot is for the levels belonging to $m_{\rm F} = 1$
and the empty box is for $m_{\rm F} = 0$.
   }
   \label{fig:gm_dnsphi}
\end{figure}

 Before going into the detailed separate discussions on the $g_s>0$ and $g_s<0$ cases, we first show in Fig.\ref{fig:gm_dnsphi}
the particle density (a) and the excitation level structure (b) for a typical example ($g_s/g_n=-0.1$ and $M/N=0.586  $) at a finite temperature ($T = 0.3 \mu K$).
As seen from Fig. \ref{fig:gm_dnsphi}(a) the condensate density with the spin +1 component in this case occupies
the central region where the confining potential is weak.
The non-condensate, which includes  all the components +1,0 and -1
evenly, distribute rather uniformly and are pushed outwardly.
The excitation level structure is shown in Fig. \ref{fig:gm_dnsphi}(b)
where the levels are plotted relative to the ground state energy level situated at $\varepsilon=0$. It is seen that 
the lowest levels come from the spin +1 component
and the other levels with the 0 and -1 components
are pushed upwards.

\subsection{Ferromagnetic interaction}

\begin{figure}
   \begin{center} \leavevmode
   \vspace{-0.8cm}

   \epsfxsize=8cm \ \epsfbox{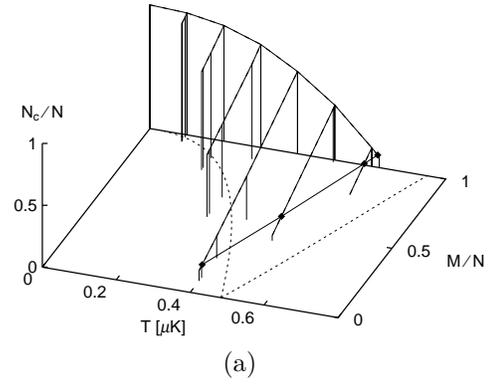} \\
   \vspace{-1\baselineskip}
   (a)\\

   \vspace{1\baselineskip}
   \epsfxsize=7cm \ \epsfbox{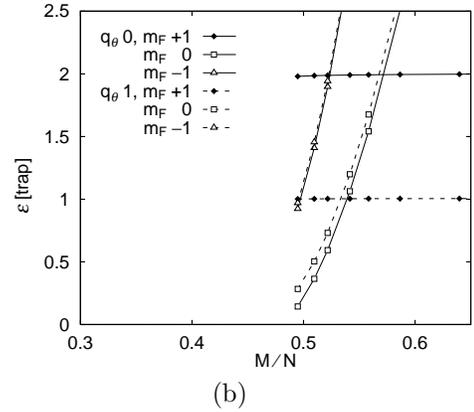} \\
   (b)\\

   \end{center}
   \caption{
   (a)  Stereographic view of the phase diagram for $g_s/g_n = - 0.1$.
   The  axes show the relative magnetization, temperature and
   the particle number of the condensate
   normalized by the total particle number.
   The dotted line on the floor is $T_{c1}$ and $T_{c2}$ lines
   of the non-interacting system.
   (b) Excitation levels as a function of $M/N$ at $T=0.3\mu K$.
The solid line means the angular momentum 
$q_{\theta} = 0$ and
the dashed line means  $q_{\theta} = 1$.
The filled dot means levels belonging to $m_{\rm F} = 1$.
The empty box means $m_{\rm F} = 0$ and
the empty triangle means $m_{\rm F} = -1$.
   } 
   \label{fig:magTcon_M}
\end{figure}

Let us start with the ferromagnetic case ($g_s<0$).
The condensate tends to have only one spin component
whose spin direction is antiparallel to the magnetic field.
Fig.~\ref{fig:magTcon_M}(a) is a stereographic view of the phase diagram where the relative number of the 
condensate is displayed. 
The curve with filled dots indicates that the particle number of the
condensate becomes 10\% of the total particle number,
signalling that the transition line on the floor is approached from low $T$.
The dotted curve on the floor is the non-interacting phase diagram shown in Fig.~\ref{fig:tc1tc2}.
It is seen that the $T_{c1}$ line is lowered due to
the interaction effect.
A drastic change from the non-interacting case occurs
at smaller $N_c/N$ and lower $T$ region where
no finite condensation solution is reached.
Therefore, upon lowering $T$ the system 
first Bose condenses and at a further lower $T$
it reenters the normal state instead of the second 
transition $T_{c2}$ as in the ideal Bose system.
In the actual situations the condensation with
$m_F = +1$  may
become phase-separated spatially into two components $m_F$=+1 and -1,
keeping the total polarization fixed.
Since the system is ferromagnetic,
the other spin components tend to be avoided.
This causes the empty region at the left hand side of the phase diagram.

Figure~\ref{fig:magTcon_M}(b)
shows the level structure of a few lowest levels.
While the levels of $m_F = +1$ denoted by the filled dots stay at the same energies,
other levels (empty symbols) belonging to $m_F = $0 and  -1 go down as the polarization $M/N$ decreases and finally reach the $\varepsilon=0$ line where the ground state level situates,
indicating that the condensation with $m_F = +1$ does not exist any further.
This lowering tendency itself agrees with the discussion for non-interacting system
in Sec.~\ref{sec:ideal}, however the latter does not exhibit the instability. This is caused solely due to the ferromagnetic interaction effect.

\subsection{Antiferromagnetic interaction}

\begin{figure}
   \begin{center} \leavevmode
   \vspace{-0.8cm}

   \epsfxsize=8cm \ \epsfbox{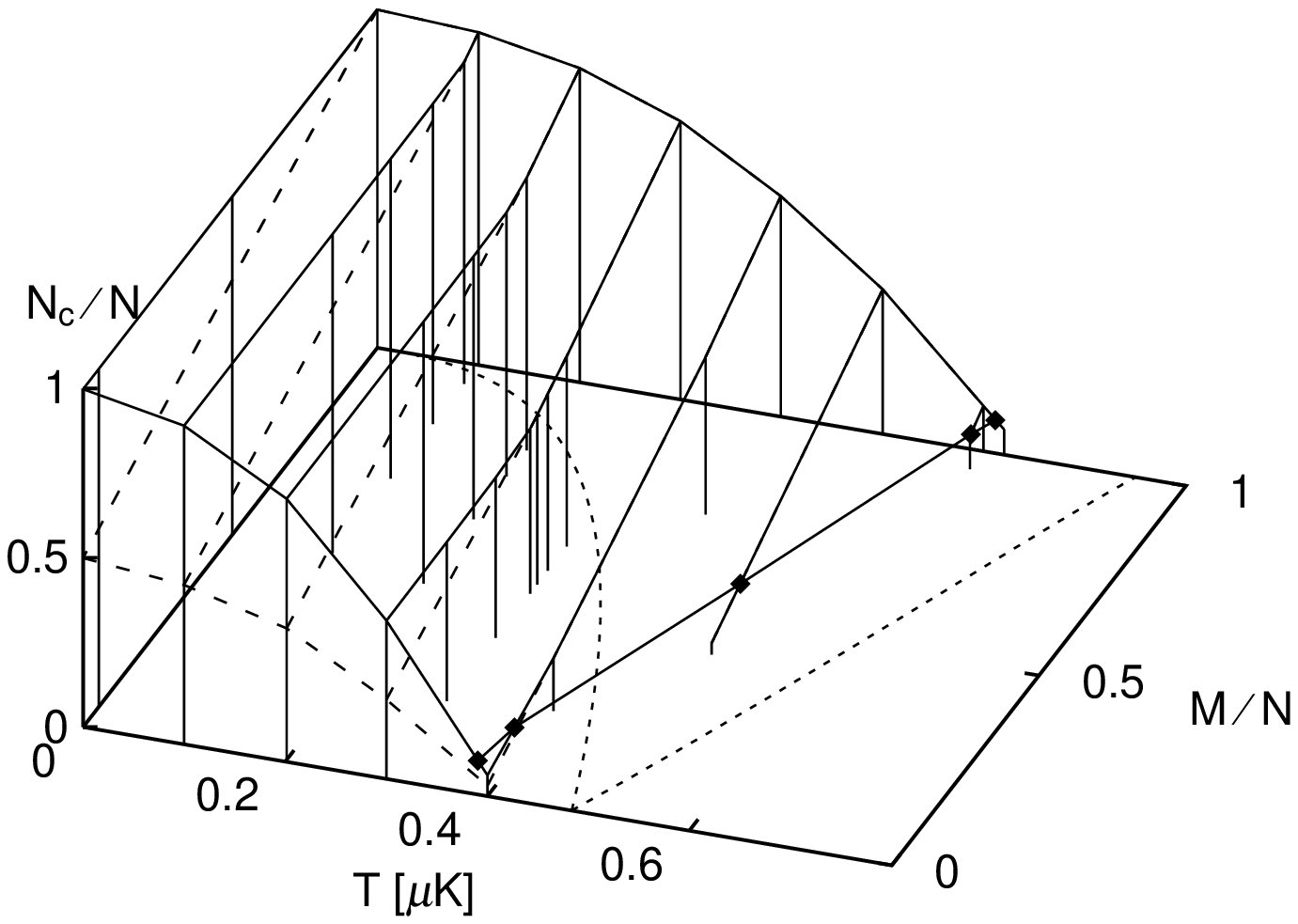} \\
   \vspace{-1\baselineskip}
   (a)\\

   \vspace{1\baselineskip}
   \epsfxsize=7cm \ \epsfbox{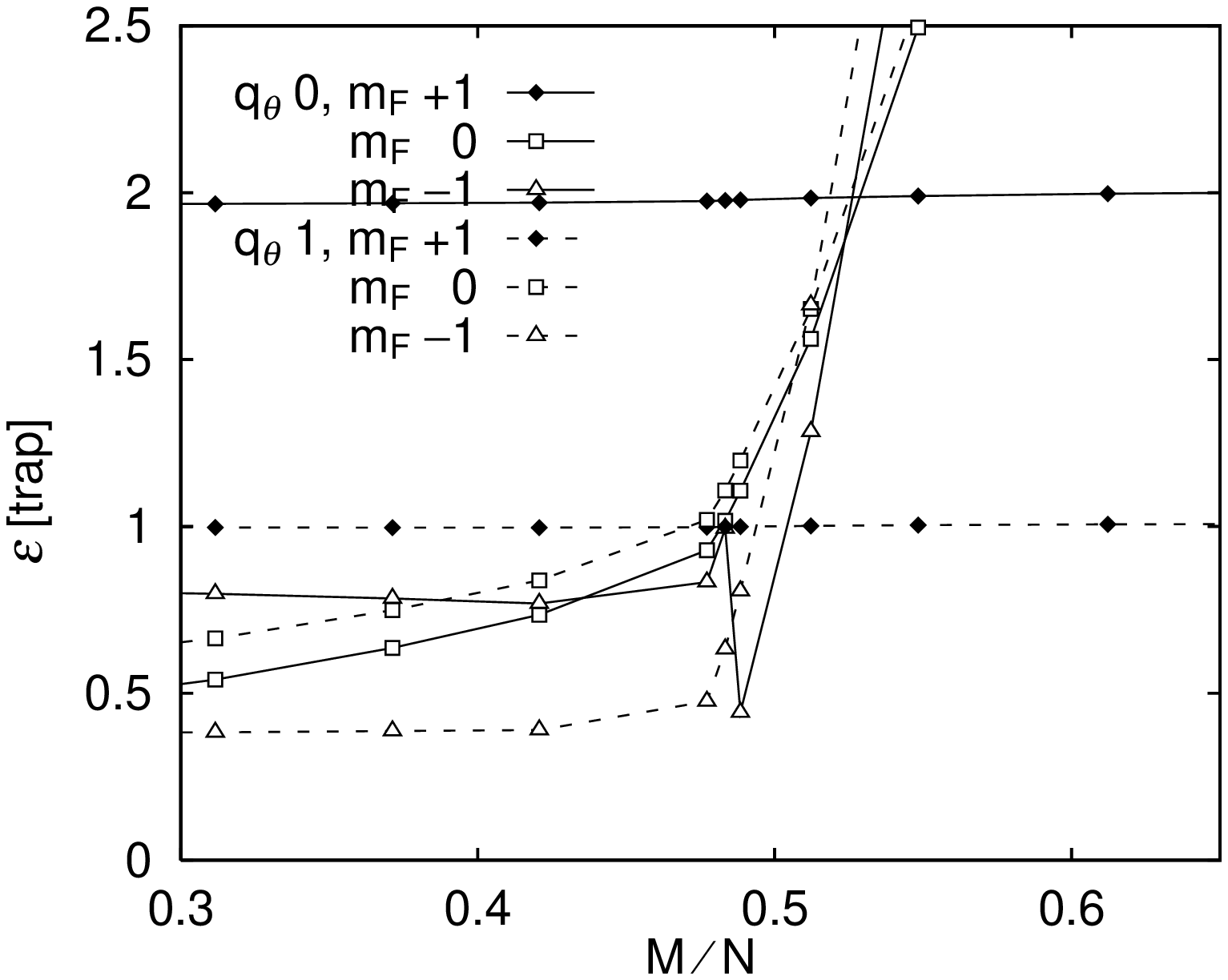} \\
   (b)\\

   \end{center}
   \caption{(a)  
   Stereographic view of the phase diagram for $g_s/g_n = + 0.1$.
  The axes show the relative magnetization, temperature and
   the particle number of the condensate
   normalized by the total particle number.
   The dotted line on the floor is $T_{c1}$ and $T_{c2}$ lines
   of the non-interacting system.
   The dashed line means the $m_F=+1$ fraction of the condensate.
   At smaller $m$, the $-1$ fraction of condensate appears (between the
   dashed line and the upper solid line).
   (b) Excitation levels as a function of $M/N$ at $T=0.3\mu K$.
The solid line means $q_{\theta} = 0$ and
the dashed line means the angular momentum $q_{\theta} = 1$.
The filled dot means levels belonging to $m_{\rm F} = 1$.
The empty box means $m_{\rm F} = 0$ and
the empty triangle means $m_{\rm F} = -1$.
   } 
   \label{fig:magTcon_P}
\end{figure}

Let us consider the antiferromagnetic case where
 $g_s$ is positive.
The same set of the parameters as in the previous Subsection except that the sign
of $g_s$ is reversed, keeping its magnitude same, gives a quite different stereographic view of the phase
diagram shown in
 Fig.~\ref{fig:magTcon_P}.
The phase diagram in the  $M/N$ vs $T$ plane
looks similar to that  in the ideal Bose case in Fig.~\ref{fig:tc1tc2}. There exist always the double transitions at $T_{c1}$ and $T_{c2}$ in the present
case. The higher transition $T_{c1}$ indicates that 
the spin component with $m_F=+1$ appears while at the second transition $T_{c2}$ the $m_F=-1$ component
appears in a continuous manner. Note that the 
$m_F=0$ condensate component never shows up.
This is a different point from the ideal Bose system
where all three components appear at lower $T$.
However, the overall behavior is essentially the same as in 
the ideal Bose case. We can see slight depressions of $T_{c1}$ and $T_{c2}$ compared to the ideal case. This is one of the interaction effects. The magnitude of the $T_c$ depression depends strongly on the density channel
interaction $g_n$, but only weakly on the spin channel
interaction $g_s$ as shown shortly.

Figure \ref{fig:magTcon_P}(b) shows the excitation level
structure at the lower energy region. 
Here the levels belonging to $m_F=-1$ and $m_F=0$
never reach the $\varepsilon=0$ line as $M/N$ decreases.
There is a zigzag of the levels at around $M/N = 0.5$.
This signals the second transition point at
$T_{c2}$, corresponding to $a=0$
in \S \ref{sec:ideal}.

In order to examine the interaction effects quantitatively,
we also calculate another set of the parameters for the system
having a weaker $g_n$ constant,
compared with the above result shown in Fig.\ \ref{fig:magTcon_P}
while keeping the ratio $g_s/g_n=0.1$ fixed.
The two transition temperatures differ from those in the  
ideal Bose case discussed in \S \ref{sec:ideal}
and also slightly from that in Fig.\ \ref{fig:magTcon_P}.
The resulting phase diagram is in between them,
namely, the two transition temperatures
become similar to that of  the ideal Bose case.

\section{Conclusion}

In this paper, we have examined the phase transition in the spinor BEC with the hyperfine state $F=1$. We focus on the interaction effects: There are two kinds of the interaction channels in the present spinor BEC, namely, the spin channel specified by $g_s$ and the density channel specified by $g_n$.
The ideal Bose system always exhibits the double transitions from the normal state 
to the one-component BEC at $T_{c1}$ and then
to the three component BEC at $T_{c2}$ upon lowering $T$.
Depending on the nature of the interaction,
or the sign of $g_s$,
only the antiferromagnetic spinor BEC ($g_s>0$) shows the 
double transitions. In this case the $m_F=-1$ component
becomes non-vanishing below the lower $T_{c2}$  where 
the $m_F=+1$ component appeared at the higher transition $T_{c1}$ persists  in the ground state.

In the ferromagnetic case ($g_s<0$), on the other hand,
the spinor
BEC always shows the instability at a lower temperature.
This lower instability transition may actually be realized as the phase-separation. The dominant $m_F=1$ component and other components segregate spatially.

The interaction $g_n$ of the density channel acts  simply to suppress both transition temperatures $T_{c1}$ and $T_{c2}$ by similar amount. Therefore it is found that 
$g_n$ modifies the system quantitatively while $g_s$
drastically alters in a qualitative manner.

We also study the spatial distribution of each spinor BEC component and the excitation spectral structure.
It is our hope that by using optical trapping method
these predictions must be checked experimentally
since we have the antiferromagnetic spinor $^{23}$Na 
BEC and $^{87}$Rb is expected to be ferromagnetic.




\end{document}